\documentclass[aps,prb,reprint,superscriptaddress,showpacs,longbibliography,
floatfix,footnoteinbib]{revtex4-1}
\usepackage{graphicx,color}
\usepackage{amsthm}
\usepackage{amsfonts}
\usepackage{algorithmic}
\usepackage{enumerate}
\usepackage{latexsym}
\usepackage{amsmath}
\usepackage{amssymb}
\usepackage[plainpages=false,pdfpagelabels,colorlinks=true,linkcolor=red,
urlcolor=blue,citecolor=blue,pdftitle={Langevin Simulations of a Long
  Range Electron Phonon Model},pdfauthor={},
pdfdisplaydoctitle=true,pdfduplex=DuplexFlipLongEdge]{hyperref}

\emergencystretch=\maxdimen
\include{MyCommand}

\begin{document}

\title{Langevin Simulations of a Long Range Electron Phonon Model}

\author{G. G. Batrouni}
\affiliation{Universit\'e C\^ote d'Azur, INPHYNI, CNRS, 0600 Nice,
  France}
\affiliation{MajuLab, CNRS-UCA-SU-NUS-NTU International Joint Research
  Unit, 117542 Singapore}
\affiliation{Centre for Quantum Technologies, National University of
  Singapore, 2 Science Drive 3, 117542 Singapore}
\affiliation{Beijing Computational Science Research Center, Beijing
  100193, China}
\author{Richard T. Scalettar}
\affiliation{Department of Physics, University of California, Davis,
California 95616, USA}

\begin{abstract}
We present a Quantum Monte Carlo (QMC) study, based on the Langevin
equation, of a Hamiltonian describing electrons coupled to phonon
degrees of freedom.  The bosonic part of the action helps control the
variation of the field in imaginary time. As a consequence, the
iterative conjugate gradient solution of the fermionic action, which
depends on the boson coordinates, converges more rapidly than in the
case of electron-electron interactions, such as the Hubbard
Hamiltonian.  Fourier Acceleration is shown to be a crucial ingredient
in reducing the equilibration and autocorrelation times.  After
describing and benchmarking the method, we present results for the
phase diagram focusing on the range of the electron-phonon 
interaction.  We delineate the regions of charge density wave
formation from those in which the fermion density is inhomogeneous,
caused by phase separation.  We show that the Langevin approach is
more efficient than the Determinant QMC method for lattice sizes $N
\gtrsim 8 \times 8$ and that it therefore opens a potential path to
problems including, for example, charge order in the 3D
Holstein model.
\end{abstract}


\maketitle

\noindent
\section{Introduction} 

Quantum Monte Carlo (QMC) constitutes one of the most powerful
non-perturbative approaches to interacting fermion Hamiltonians.  Its
applications have led to insight into both renormalized single
particle properties at low density, and also to many-body phase
transitions in systems ranging from high energy, to nuclear, to
condensed matter physics.  Nevertheless, fermion QMC suffers from
several serious limitations: (i) Algorithms generally scale as the
cube of the number of particles or sites, a consequence of the
evaluation of the determinants arising when the fermionic degrees of
freedom are integrated out.  (ii) Depending on the model and on the
parameter regime, long autocorrelation times can also challenge the
calculations.  Finally, (iii) the fermion sign problem remains the
most restrictive bottleneck in the field.

Considerable progress has been made in addressing (i) in the lattice
gauge theory (LGT) community, {\it e.g.} with linear scaling methods
for the fermion determinant evaluation.  However, these techniques
have proven surprisingly difficult to carry over into condensed matter
(CM) and specifically to the Hubbard model
\cite{scalettar86,scalettar87,yamazaki09,bai09,beyl18}.  The general
belief is that the difficulty arises from two related problems: the
high degree of anisotropy between imaginary time ($\beta$) and the
spatial dimensions present in CM problems, as opposed to the
relativistic LGT case, and the much more rapid variation of the
Hubbard-Stratonovich field configurations in the imaginary time
direction which gives rise to less well-conditioned matrices.  The
eigenvalue spread in CM problems leads to large conjugate gradient
iteration counts, and, indeed, often, to a complete absence of
convergence of iterative matrix inversion solvers.

In this paper we explore the application of linear scaling QMC methods
to electron-phonon Hamiltonians.  We are motivated by the fact that,
in such models, the kinetic energy terms $\sum_i \hat p_i^2/2$ in the
oscillator Hamiltonian smooth the imaginary time variation of the
phonon field.  As a consequence, the condition number of the fermion
determinants is likely to be improved relative to that which arises in
the Hubbard model, where the Hubbard-Stratonovich field has no similar
dynamics.  It is for the latter case that most of the linear scaling
methods have previously been tested
\cite{scalettar86,scalettar87,yamazaki09,bai09} and shown to be
effective only in a limited parameter regime of weak coupling and/or
relatively high temperature.

The successful application of Langevin methods to electron-phonon
models would be a considerable step forward, since their physics holds
significant intrinsic interest.  Early QMC work focused on the dilute
limit.  As an electron moves through a material, the polarization of
the underlying medium causes a cloud of phonons to follow.
Simulations studied the resulting ``single electron polaron'',
identifying its size and effective mass as functions of the
electron-phonon coupling and phonon frequency
\cite{kornilovitch98,kornilovitch99,alexandrov00,hohenadler04,ku02,spencer05,macridin04,romero99,bonca99,chandler14}.
If the interaction $\lambda$ is sufficiently large, two polarons can
pair with a bipolaron size and dispersion which depend on
$\lambda$\cite{hague09}.

Meanwhile, as the density increases, QMC has explored how local up and
down spin pairs, which form due the effective interaction mediated by
the lattice distortion, can arrange themselves spatially into charge
density wave (CDW) patterns, especially on bipartite
lattices\cite{scalettar89,noack91,niyaz93}.  The critical temperatures
of transitions to long range order phases have been
evaluated\cite{weber17,costa18}. These CDW patterns typically induce
insulating behavior, and compete with superconducting phases which can
also occur if the pairs become phase coherent across the
lattice\cite{scalettar89,costa18}.

In the remainder of this paper, Sec.~2 will describe the specific
Hamiltonian to be studied, its experimental motivation, and the
details of the Langevin method.  Section 3 contains benchmarks of our
results and comparisons with other methods.  Section 4 describes
results for the low temperature ordered phases of the Hamiltonian, and
Sec.~5 presents some concluding remarks.

\noindent
\section{Model and Methods} 

We will study the properties of a two-dimensional system governed by
the Hamiltonian,
\begin{align}
\hat H &= \hat H^0_{\rm el} + \hat H^0_{\rm ph} + \hat V_{\rm el-ph},
\label{ham}\\
\hat H^0_{\rm el} &= -t \sum_{\langle ij \rangle \sigma}
\big( \, \hat c_{i\sigma}^{\dagger} \hat c_{j\sigma}^{\phantom{\dagger}}
+ \hat c_{j\sigma}^{\dagger} \hat c_{i\sigma}^{\phantom{\dagger}} \,
\big) -\mu \sum_{i\sigma} \hat n^{\phantom{\dagger}}_{i\sigma},
\nonumber \\
\hat H^0_{\rm ph} &= \frac{1}{2} \, \omega_0^2 \, \sum_i \hat x_i^2
+ \frac{1}{2} \, \sum_i \, \hat p_i^2 \,\,,
\nonumber\\
\nonumber
\hat V_{\rm el-ph} &=  \lambda_0 \sum_{ir\sigma}
f(r) \, \hat x_i  \, \hat n_{i + r,\sigma},
\hskip0.22in
f(r) = \frac{e^{-r/\xi}}{(1+r^2)^{3/2} }.
\end{align}
$\hat c_{i\sigma}^{\dagger}$ and $\hat
c_{j\sigma}^{\phantom{\dagger}}$ are creation and destruction
operators for electrons of spin $\sigma$ on lattice site $i$, so that
$\hat H^0_{\rm el}$ describes the hopping of electrons of spin
$\sigma$ between near neighbor sites $\langle ij\rangle$.  Here we
focus on a 2D square lattice and choose $t=1$ to set the energy scale.
$\hat H^0_{\rm ph}$ represents a dispersionless (optical) phonon mode
on each lattice site.  In $\hat V_{\rm el-ph}$, the phonon
displacement $\hat x_i$ on site $i$ couples to the electron densities
$\hat n_{i+r,\sigma}$ with a strength $\lambda_0 f(r)$ which falls off
exponentially with separation $r$\cite{frohlich54}.  Our simulations
are done in the grand-canonical ensemble, with a chemical potential
$\mu$ which couples to the fermion density $\hat
n^{\phantom{\dagger}}_{i\sigma} = \hat c^{\dagger}_{i\sigma} \hat
c^{\phantom{\dagger}}_{i\sigma}$.  The model has a particle-hole
symmetry on a bipartite lattice; half-filling is attained at $\mu = -
(\sum_r \lambda(r))^2/\omega_0^2$.  In the literature, $\hat V_{\rm
  el-ph}$ is sometimes written by replacing $\hat x_i \rightarrow (\,
\hat a^{\dagger}_i + \hat a^{\phantom{\dagger}}_i \,) / \sqrt{2
  \omega_0}$ and defining the coupling $g=\lambda_0 /
\sqrt{2\omega_0}$.  We will use this notation in our work as well.

The Holstein model\cite{holstein59}, which has been the focus of most
of the previous QMC investigations, is obtained in the extreme
short-range limit $\xi \rightarrow 0$ where the phonon mode on site
$i$ couples only to the electron density on the same site.  Since the
Holstein electron-phonon coupling is absolutely local in space, it has
no momentum dependence.  Equation \ref{ham}, on the other hand, allows
for (a specific) $\tilde \lambda(q)$ via the Fourier transform of
$\lambda_0 f(r)$.  Continuous Time Quantum Monte Carlo (CTQMC) has
been used to study the effect of varying the range $\xi$ on polaron
and bipolaron formation\cite{hague09}.  A momentum averaging technique
has also been developed to study general $\lambda(q)$
\cite{goodvin08}.

The interest in studying momentum-dependent coupling is driven by a
number of factors.  First, a momentum dependent $\tilde \lambda(q)$
has been suggested to play a role in superconducting transitions in
FeSe monolayers\cite{wang16}, electron-phonon physics in SrTiO$_3$
\cite{swartz16}, and the behavior of 2H-NbSe$_2$ \cite{weber11}.
Second, there are qualitative issues to be addressed, {\it e.g.}~how
the range of the electron phonon interaction, $\xi$, affects the
competition between metallic and Peierls/CDW phases at half-filling.
Here, recent CTQMC studies in one dimension have shown that as $\xi$
increases from zero, the metallic phase is stabilized and, for
sufficiently large $\lambda_0$, phase separation can also occur
\cite{hohenadler12}.  Finally, it has been suggested that
materials-specific forms for $\lambda(q)$ can be incorporated into QMC
simulations of appropriate model Hamiltonians \cite{devereaux16}.

The solution of Eq.(\ref{ham}) via QMC proceeds as follows.  The
inverse temperature $\beta$ is discretized into $L_\tau$ intervals of
length $\Delta \tau \equiv \beta/L_{\tau}$.  Complete sets of phonon
variables $\{x(i,\tau),p(i,\tau)\}$ are introduced at each imaginary
time slice of the partition function ${\cal Z} = {\rm Tr} \, e^{-\beta
  \hat H}= {\rm Tr} \, e^{-\Delta \tau \hat H} e^{-\Delta \tau \hat H}
\cdots e^{-\Delta \tau \hat H}$.  We integrate out the momentum and,
since the Hamiltonian is quadratic in the fermionic operators, they
can be traced out finally giving\cite{scalettar89},
\begin{align}
\label{partition1}
{\cal Z}&= \int {\cal D} \, x(i,\tau) \,\, e^{- S_{\rm bose} } \,\,
\big(\, {\rm det} \, M(\{x(i,\tau)\}\, \big)^2, \\ 
\nonumber
S_{\rm bose} &= \frac{1}{2} \Delta\tau \omega^2 \sum_{i,\tau}
x(i,\tau)^2 \\
\nonumber
& + \frac{1}{2} \Delta\tau\sum_{i,\tau} \Big( \, \frac{x(i,\tau+1) -
  x(i,\tau)}{\Delta \tau}\Big)^2, \nonumber
\end{align}
is an integral over the space and imaginary time dependent scalar
phonon field $x(i,\tau)$.  The integrand has a ``bosonic'' piece,
$S_{\rm bose}$, originating in $\hat H_{\rm ph}^0$ and {\it identical}
fermion determinants (one for each spin species) arising from
integrating out the fermions.  The matrix elements of $M$ depend on
the phonon field.  Details of the form of $M$ are in
Ref.~[\onlinecite{blankenbecler81}].  This general approach, is often
referred to as the ``Determinant Quantum Monte Carlo'' (DQMC)
method\cite{blankenbecler81,sorella89}.

DQMC can be applied to models with el-el (as opposed to el-ph)
interactions like the Hubbard Hamiltonian via the introduction of a
Hubbard-Stratonivich (HS) field which decouples the quartic
interaction terms to quadratics, allowing the fermion trace to be
performed as above.  In such applications, there is typically a sign
problem- the product ${\rm det}\, M_\uparrow(\{x(i,\tau)\}) \,\,
{\rm}\, {\rm det} M_\downarrow(\{x(i,\tau)\})$ can become negative,
precluding the sampling of the HS Field.  A crucial observation for
the electron-phonon model of Eq.~1 is that the up and down fermion
matrices are identical (hence their spin index is suppressed in
Eq.~2.), and there is no sign problem.

At this point there are two approaches.  In almost all CM
applications, the determinant of $M$, which is a very sparse dimension
$L_\tau N$ matrix, is rewritten as the determinant of a smaller,
dense, dimension $N$ matrix where $N$ is the number of sites.  Changes
to the phonon field variables $x(i,\tau)$ are performed individually
and, because of their local nature, the change in ${\rm det} \, M$ can
be evaluated in ${\cal O}(1)$ operations if the Green function $G =
M^{-1}$ is known.  After each change, $G$ must be updated, a process
which takes ${\cal O}(N^2)$ steps, since the change to a single
$x(i,\tau)$ involves only a rank one alteration of $M$.  A sweep of
all $N L_\tau$ variables then requires ${\cal O}(N^3 L_\tau)$ steps.

The second approach (used in the majority of LGT applications) retains
the larger sparse matrix $M$ as the central object.  All degrees of
freedom are updated simultaneously using the Langevin equation, in a
manner that is linear in both $N$ and $L_\tau$, under the assumption
that the sparse linear algebra solver does not have an iteration count
which increases with system size.  This assumption fails dramatically
for the Hubbard model, where there is no $S_{\rm bose}$ for the HS
field.  We will explore here the efficacy of the Langevin approach for
electron-phonon models.  Many subtleties need to be carefully assessed
to perform a meaningful comparison with the ${\cal O}(N^3 L_\tau)$
approach.  First, for a choice of Hamiltonian parameters, the
dependence of the iteration count on $N$ and $L_\tau$ must be
monitored.  Second, the equilibration and autocorrelation times must
be measured.  At a minimum, the iteration count and correlation times
provide a potentially large prefactor to the linear scaling, competing
with the savings due to linear scaling as opposed to cubic scaling.
Finally, the effect of the discretization of the Langevin evolution on
physical observables must be determined.  These issues must be well
understood as a function of the parameters in the Hamiltonian (the
location in phase space).

We now describe the details of our approach, which is based on the
algorithm in Ref.~[\onlinecite{batrouni85}]. We first write
Eq.~(\ref{partition1}) in the form,
\begin{equation}
{\cal Z} = \int {\cal D} \, x(i,\tau) \,\, e^{- S},
\label{partition2}
\end{equation}
where
\begin{equation}
S = S_{\rm bose} - {\rm ln}({\rm det}M)^2,
\label{action1}
\end{equation}
and define fictitious dynamics governed by the Langevin equation,
\begin{equation}
\frac{{\rm d}x(j,\tau,t)}{{\rm d}t}=-\frac{\partial
  S}{\partial x(j,\tau,t)} + \sqrt{2}\, \eta(j,\tau,t),
\label{langevin1}
\end{equation}
with the stochastic variable $\eta$ satisfying
\begin{equation}
\langle \eta(j,\tau,t)\rangle =0,\, \langle
\eta(j,\tau,t) \eta(r,\tau^\prime,t^\prime)\rangle
=\delta_{j,r} \delta_{\tau,\tau^\prime} \delta(t-t^\prime).
\label{gaussian}
\end{equation}
In Eqs.~(\ref{langevin1},\ref{gaussian}), $j$ labels the spatial
coordinate of a site, $\tau$ its imaginary time and $t$ the Langevin
time. The condition given by Eq.~(\ref{gaussian}) can be satisfied by
taking the stochastic variables $\eta$ to be random numbers with
Gaussian distribution. The stationary limit of the statistical weight
of the configurations, $P(\{x(i,\tau)\})$, can be determined by first
writing the Fokker-Planck equation associated with
Eq.~(\ref{langevin1}) which describes the time evolution of $P$. It
can then be easily shown\cite{batrouni85} that in the long time limit
the distribution is given by $P={\rm exp}(-S)$, justifying using
Eq.~(\ref{langevin1}) to generate configurations which are used to
calculate physical quantities.

To integrate Eq.~(\ref{langevin1}), we first discretize the Langevin
time. The simplest, Euler, discretization leads to\cite{batrouni85},
\begin{eqnarray}
\nonumber
x(j,\tau,t+{\rm d}t)&=&x(j,\tau,t) - {\rm d}t \frac{\partial
  S}{\partial x(j,\tau,t)} \\
&&+ \sqrt{2{\rm d}t}\, \eta(j,\tau,t),
\label{euler}
\end{eqnarray}
with 
\begin{eqnarray}
\nonumber
\langle \eta(j,\tau,t\rangle&=&0,\\
\langle\eta(j,\tau,t)\eta(r,\tau^\prime,t^\prime)\rangle &=& 
\delta_{j,r^\prime} \delta_{\tau,\tau^\prime}\delta_{t,t^\prime}.
\label{eulernoise}
\end{eqnarray}
Note the square root of the Langevin time step, ${\rm d}t$, in
Eq.~(\ref{euler}) which comes from replacing the Dirac
$\delta$-function in Eq.~(\ref{gaussian}) by the discrete Kronecker
$\delta$, $\delta(t-t^\prime) \to \delta_{t,t^\prime}/{\rm
  d}t$. Because of this $\sqrt{{\rm d}t}$, the Euler discretization
error of this stochastic differential equation is ${\cal O}({\rm d}t)$
(for ${\rm d}t$ small enough) instead of the ${\cal O}({\rm d}t^2)$
for deterministic differential equations. To reduce the error in
Eq.~(\ref{euler}) to ${\cal O}({\rm d}t^2)$, one can use Runge-Kutta
discretizations adapted to stochastic differential
equations\cite{batrouni85}. However, in this work, we have found that
the simple Euler discretization has sufficient precision.

Now we deal with the action term,
\begin{align}
\nonumber
\frac{\partial S}{\partial x(j,\tau,t)} &= \frac{S_{\rm
    bose}}{\partial x(j,\tau,t)} - \frac{\partial \, {\rm ln}({\rm
    det}M)^2}{\partial x(j,\tau,t)},\\ 
&=\frac{\partial S_{\rm bose}}{\partial x(j,\ell,t)} - 2\,{\rm Tr}\left (
\frac{\partial M}{\partial x(j,\ell,t)} M^{-1}  \right ).
\label{dsdx}
\end{align}
The trace term in Eq.~(\ref{dsdx}) is expensive due to $M^{-1}$;
calculating the inverse of a matrix scales as the cube of its
dimension. In order to avoid this, we note that, given a vector of
Gaussian random numbers, $\vec g$, and a matrix $A$, we have $\langle
\, {\vec g}^{\,T} A \, {\vec g} \, \rangle = {\rm Tr} A$ where the
average is taken over the Gaussian distribution of the random
numbers. This allows us to replace the trace term in Eq.~(\ref{dsdx})
by a stochastic estimator,
\begin{equation}
2 \, {\rm Tr}\left ( \frac{\partial M}{\partial x(j,\ell,t)}
M^{-1}  \right ) \Longrightarrow  2 \, {\vec g}^{\,T}  \left (
\frac{\partial M}{\partial x(j,\ell,t)} M^{-1}  \right )
     {\vec g}.
\label{estimator}
\end{equation}
We recall here that we are using the large sparse form of the matrix
$M$, {\it i.e.} a sparse matrix with dimension $NL_\tau$.
Consequently, $\vec g$ is a vector with $NL_\tau$ elements. Using this
estimator, we avoid having to calculate $M^{-1}$ because what is
needed now is $M^{-1}\vec g$, which is much faster to evaluate using,
for example, the bi-Conjugate Gradient (CG) algorithm to solve $M
{\vec v}=\vec g$. For a positive matrix, CG is guaranteed to converge
to the exact result in at most $NL_\tau$ iterations.  The issue of the
positivity of $M$ is a subtle one which we shall not address
here\cite{scalettar86,scalettar87,yamazaki09,bai09}.  Of course, one
does not need the exact answer and, instead, sets a precision
threshold at which the CG iterations are stopped. Usually this leads
to a rather small number of iterations (see below).

The Langevin iterations are implemented using
Eqs.~(\ref{euler},\ref{gaussian}) with
\begin{equation}
\frac{\partial S}{\partial x(j,\tau,t)} = \frac{S_{\rm bose}}{\partial
  x(j,\tau,t)} - 2{\vec g}^{\,T} \left ( \frac{\partial M}{\partial
  \phi({\bf j},\ell,t)} M^{-1} \right ) {\vec g}.
\label{dsdxestimator}
\end{equation}
Note that in this algorithm, the entire phonon field, ${x(j,\tau)}$,
is updated in a single step. All the operations are simple sparse
matrix-vector multiplies which can be easily optimized.

One of the main problems, mentioned earlier, facing simulations of
electron-phonon systems is the very long auto-correlation times.  We
introduce Fourier acceleration (FA) which helps reduce this
problem. We first note that Eq.~(\ref{langevin1}) is just one in an
infinite class of Langevin equations, all of which lead to the same
stationary limit. Consider an arbitrary but positive definite matrix
$Q$.  Configurations generated by the Langevin equation,
\begin{equation}
\frac{{\rm d}{\vec x}(t)}{{\rm d}t} = - Q \frac{{\rm d}S}{{\rm d}{\vec
    x(t)}} + \sqrt{2Q}\,{\vec \eta}(t),
\end{equation}
are guaranteed to be given by the correct distribution (${\rm
  exp}(-S)$) in the long time limit regardless of the form of
$Q$. This additional flexibility offers the possibility of choosing
$Q$ to shorten autocorrelation times, leading to accelerated
convergence. To guide our choice, we note that in the noninteracting
limit, $\lambda_0=0$, we have,
\begin{eqnarray}
\nonumber \frac{{\rm d}S}{{\rm d}x(i,\tau)}&=&\Delta\tau \, \omega^2
x(i,\tau)\\ &&
+\frac{[x(i,\tau+1)+x(i,\tau-1)-2x(i,\tau)]}{\Delta\tau},
\label{freedsdx}
\end{eqnarray}
which becomes, after Fourier transforming along imaginary time,
\begin{equation}
\frac{{\rm d}{\tilde S}}{{\rm d}{\tilde x}(i,k_\tau)}=\left
(\Delta\tau \, \omega_0^2 +[2-2\, {\rm cos}(2\pi k_\tau/L_\tau)] /\Delta
\tau \right ){\tilde x}(i,k_\tau),
\label{freemom}
\end{equation}
with $-L_\tau/2 + 1\leq k_\tau \leq L_\tau/2$. We see that the ratio
of the slowest to fastest mode is,
\begin{equation}
\frac{(\Delta \tau \, \omega_0)^2}{4+(\Delta \tau \, \omega_0)^2} \ll 1,
\label{critslow}
\end{equation}
exposing the critical slowing down of the phonons in the imaginary
time direction, especially at small $\Delta \tau$. To compensate for
this, we choose the matrix $Q$ to be diagonal in imaginary time
Fourier space and given by,
\begin{equation}
{\tilde Q}(k_\tau) = \frac{\Delta \tau\omega_0^2 + 4/ \Delta \tau}{
    \Delta \tau\omega_0^2+(2-2\, {\rm cos}(2\pi k_\tau/L_\tau))/ \Delta\tau},
\label{accmat}
\end{equation}
which is normalized so that ${\tilde Q}(L_{\tau}/2)=1$. In the
noninteracting limit, this choice will totally eliminate critical
slowing down. This is clearly not true when $\lambda_0 \neq
0$. Nonetheless we find that this form, motivated by the
noninteracting limit, works very well and helps convergence even in
the strongly interacting case. We, therefore, use this form in all the
follows.

Our Langevin equation now becomes,
\begin{eqnarray}
\nonumber
\frac{{\rm d}{\vec x}(t)}{{\rm d}t} &=& - {\bf \hat F}^{-1}{\tilde
  Q}(k_\tau){\bf \hat F} \frac{{\rm d}S}{{\rm d}{\vec x(t)}} +
     {\bf\hat F}^{-1}\sqrt{2{\tilde Q(k_\tau)}}\,{\bf\hat F}{\vec
       \eta}(t)\\
&=& - {\bf \hat F}^{-1}\bigg [
{\tilde Q}(k_\tau){\bf \hat F} \frac{{\rm d}S}{{\rm d}{\vec x(t)}}
+ \sqrt{2{\tilde Q(k_\tau)}}\,{\bf\hat F}{\vec \eta}(t) \bigg ],
\label{acclangevin}
\end{eqnarray}
where ${\bf \hat F}$ is an FFT operator, $\tilde Q(k_\tau)$ is given
by Eq.~(\ref{accmat}), and ${\rm d}S/{\rm d}{\vec x}$ is given by
Eq.~(\ref{dsdxestimator}).

Calculating phonon quantities is straight forward since the QMC
evolves the phonon field directly. All fermioinc quantities can be
calculated once the Green function is obtained. This is given by,
\begin{equation}
G(i,j) = \left \langle \left (M[\{x\}]\right )_{i,j}^{-1}\right
\rangle,
\label{green1}
\end{equation}
where the sites $i$ and $j$ can be at equal or unequal imaginary time.
(Indeed, an additional advantage of the Langevin approach is that one
does not need separate, and computationally costly, routines to
evaluate the unequal time Green function.)  As we did in the update
steps, we avoid evaluating the inverse of the matrix $M$ by
calculating the Green function using a stochastic estimator,
\begin{equation}
G(i,j) = \left \langle \gamma_i \left (M[\{x\}]^{-1}\vec \gamma\right
)_j \right \rangle,
\label{green2}
\end{equation}
where $\gamma_j$ is a Gaussian random number. $M^{-1}\vec
\gamma$ is calculated with the CG algorithm. Once $G(i,j)$ is
calculated, all fermionic quantities, {\it e.g.} kinetic energy,
density correlations, structure factor etc, can be obtained.

Long range CDW order is identified by studying $S(\pi,\pi)$ where the
structure factor is given by
\begin{equation}
S(k_x,k_y) = \frac{1}{N}\sum_{{\vec r}} {\rm e}^{i {\vec k}\cdot
  {\vec r}} \langle n(0)n(\vec r)\rangle.
\label{structfact}
\end{equation}

\noindent
\section{Benchmarking the Algorithm} 

We begin by addressing the first of the two themes of this paper: an
investigation of algorithmic efficiency of Langevin-based linear
scaling (iterative) methods in electron-phonon models.

\begin{figure}[!ht]
\centerline{\includegraphics[width=9cm]{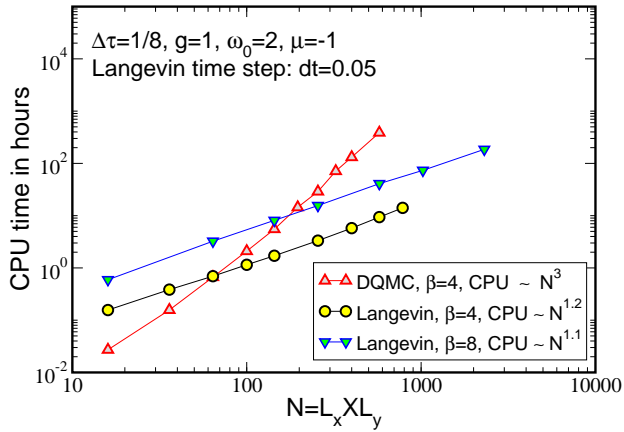}}
\caption{(Color online) CPU time versus system size in the Holstein
  limit ($\xi \to 0$) for DQMC and Langevin algorithms. DQMC scales as
  $N^3$ while Langevin scales essentially linearly for both values of
  $\beta$. $\lambda_0=g\sqrt{2\omega_0}$. }
\label{timescaling}
\end{figure}

Figure \ref{timescaling} shows the of CPU time for the Langevin and
DQMC algorithms as a function of system size $N=L_x \times L_y$ at
phonon frequency $\omega_0=2$, electron phonon coupling $g=1$ and $\xi
\to 0$ ({\it i.e.}~the Holstein contact interaction limit).  The
chemical potential $\mu=-1$ is set so that the lattice is half-filled,
$\rho=1$. For this figure, the same number of sweeps was chosen for
all the DQMC sizes, and a different fixed number of sweeps for the
Langevin runs. The goal is simply to show the scaling of CPU time.

The log-log plot demonstrates the expected $N^3$ scaling for DQMC, and
a near-linear scaling for the Langevin approach.  That the power is
slightly larger than one is a consequence of a modest increase in the
number of conjugate gradient iterations with $N$.  The Langevin
approach already becomes more efficient than DQMC for relatively small
lattice sizes, $N \sim 8 \times 8$ for $\beta=4$.  A key feature of
Fig.~\ref{timescaling} is that the (near) linear scaling is no worse
at $\beta=8$ than at $\beta=4$.  In contrast, in simulations of the
Hubbard model, the number of conjugate gradient iterations grows very
rapidly at large $\beta$ and strong coupling $U$.

\begin{figure}[!ht]
\centerline{\includegraphics[width=9cm]{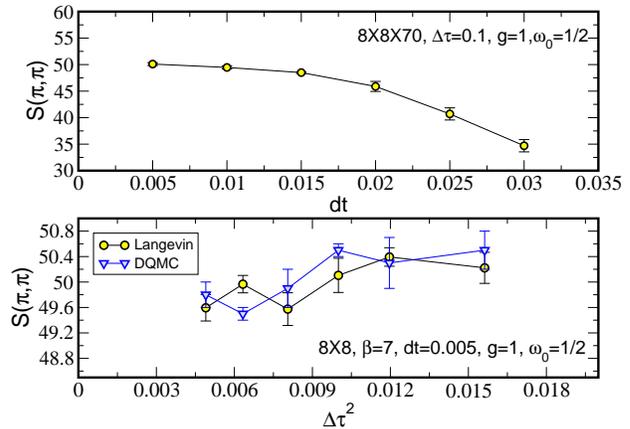}}
\caption{(Color online) Top: The dependence of $S(\pi,\pi)$ on the
  Langevin time step, ${\rm d}t$ in the ordered phase.  For ${\rm d}t
  \leq 0.01$ the discretization error becomes linear in ${\rm
    d}t$. Bottom: The dependence of $S(\pi,\pi)$ on the imaginary time
  step, $\Delta \tau$, for both DQMC and Langevin.  The Trotter-Suzuki
  errors are comparable in the two algorithms.  ${\rm d}t=0.005$ used
  to integrate the Langevin equation gives excellent agreement with
  DQMC. The data are for the Holstein limit ($\xi\to 0$).}
\label{dtdtauscaling}
\end{figure}

Figure \ref{dtdtauscaling} addresses the systematic errors in the
Langevin approach.  In the top panel, the charge density structure
factor is shown as a function of Langevin step ${\rm d}t$.  The
lattice size is $N=8\times 8$ and the inverse temperature $\beta=7$
(deep in the ordered phase) with $\Delta \tau =0.1$. It is seen that
for ${\rm d}t$ less than around $0.015$, the value of the structure
factor changes very little and, in fact, the error varies linearly
with ${\rm d}t$. For ${\rm d}t> 0.015$, the error increases rapidly
until the iterative process is destabilized beyond ${\rm d}t\approx
0.3$ for the parameters dshown in the figure. In the bottom panel the
Trotter errors are assessed at fixed Langevin step ${\rm d}t=0.005$
and compared with those arising from DQMC.  They are seen to be less
than one percent up to $\Delta\tau^2 = 0.016$ ($\Delta \tau=0.125$),
and are comparable for the two methods.

\begin{figure}[!ht]
\centerline{\includegraphics[width=9cm]{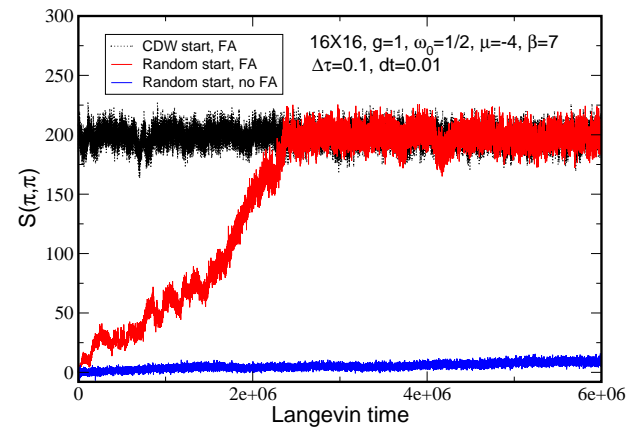}}
\caption{(Color online) $S(\pi,\pi)$ as a function of Langevin time,
  comparing Fourier acceleration (FA) with unaccelerated
  evolution. The red (FA) and blue (no FA) curves have the same random
  initial configuration of the phonon field. It is clear that for
  large systems FA is crucial for equilibration. The initial
  configuration for the black curve is CDW: The phonon field has an
  initial checkerboard configuration. }
\label{scdwvsiter}
\end{figure}

Equilibration and autocorrelation times play a key role in the
assessment of any algorithm.  Figure \ref{scdwvsiter} shows the
equilibration of the charge density structure factor $S(\pi,\pi)$ on a
$N=16 \times 16$ lattice at $\beta=7$ with $\omega_0=0.5$, $g=1$
($\lambda_0=1)$ and $\xi \to 0$.  The lattice is half-filled
($\mu=-4$).  In the absence of Fourier Acceleration, and with a random
start, the system remains in the disordered state ($S(\pi,\pi)$ small)
even out to six million time steps.  On the other hand, if FA is
implemented, it grows steadily, achieving a value consistent with the
known CDW order at $t \sim 2\times 10^6$.  If the simulation is
started in a CDW pattern, it remains in that phase.

\begin{figure}[!ht]
\centerline{\includegraphics[width=9cm]{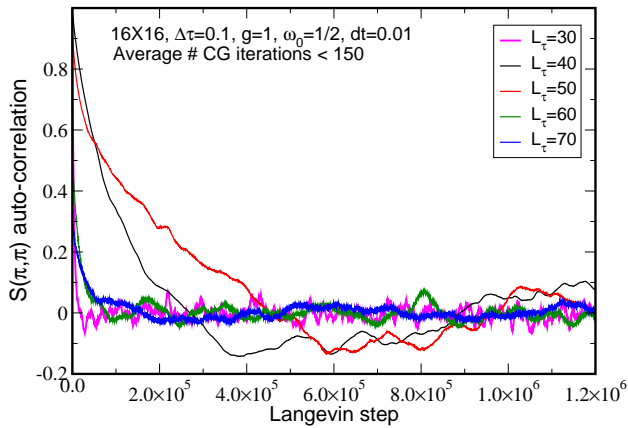}}
\caption{(Color online) The autocorrelation function of accelerated
  Langevin dynamics of the Holstein model for several $\beta$
  values. The relaxation time is longest for $L_\tau=40,\,50$
  ($\beta=4,\,5$) near the critical temperature, $\beta\approx 4.5$.}
\label{autocorr}
\end{figure}

Figure \ref{autocorr} shows the autocorrelation function in the
presence of FA for a $N=16 \times 16$ lattice.  Deep in the CDW phase,
$L_\tau=60,70$ ($\beta=6, 7$), as well as in the high temperature
phase, $L_\tau = 30$ ($\beta = 3$) the autocorrelation time is
relatively short.  As is typical, autocorrelation times are long near
the critical $\beta_c \sim 4.5$ as seen from the data with $L_\tau
=40, 50$ ($\beta = 4, 5$).

\section{The Phase Diagram of the Long Range Model} 

We focus in this paper on a longer range el-ph coupling given by
Eq.~1.  Several initial efforts have been made to study this
situation\cite{johnston18,devereaux18}. They have found a significant
tendency to phase separation.  The qualitative physics behind this is
clear: In the Holstein model an electron of one spin distorts the
phonon on its same site, attracting an electron of opposite spin
there.  The Pauli principle precludes any further clustering.  If,
however, the interaction extends to neighboring sites, electrons will
get attracted there.  These, in turn will bring in yet more particles
on next-near neighbor sites.  This cascading effect costs kinetic
energy, and entropy, but still might dominate the physics.  It has
proven difficult to expose the extent of phase separation in
traditional DQMC algorithms, since the longer range interaction makes
the update of the Green function more expensive.  At minimum there is
a succession of rank one updates whose number equals the number of
sites within the interaction range.  If $\xi$ is large enough, the
expense goes from ${\cal O}(N^3)$ to ${\cal O}(N^4)$.

Our specific goal is to get the ground state phase diagram in the
$\lambda_0-\xi$ plane.  We begin by studying the real space
density-density correlation function and its Fourier transform, the
CDW structure factor, Eq.(\ref{structfact}).  In the disordered phase,
$S(\pi,\pi)$ picks up contributions only from a small number of terms,
while in the ordered phase, $S(\pi,\pi)$ will grow linearly with $N$.

\begin{figure}[!ht]
\centerline{\includegraphics[width=9cm]{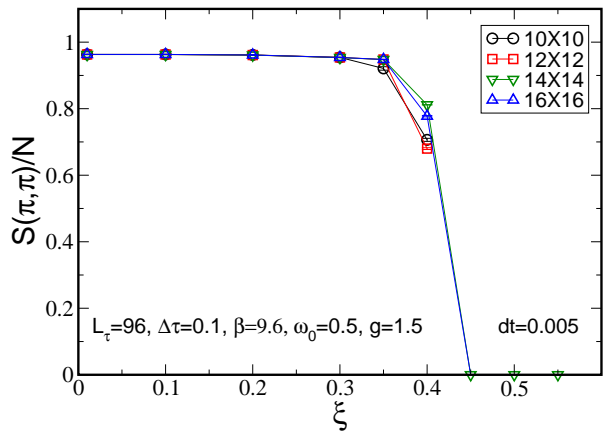}}
\caption{(Color online) The structure factor, $S(\pi,\pi)$ as a
  function of the range of the interaction, $\xi$, keeping the chemical
  potential tuned to half filling. When the interaction is large
  enough, $S(\pi,\pi)$ drops to zero due to phase separation. See
the phase diagram of
  Fig.~\ref{phasediag}.}
\label{scdwvsxi}
\end{figure}

\begin{figure}[!ht]
\centerline{\includegraphics[width=9cm]{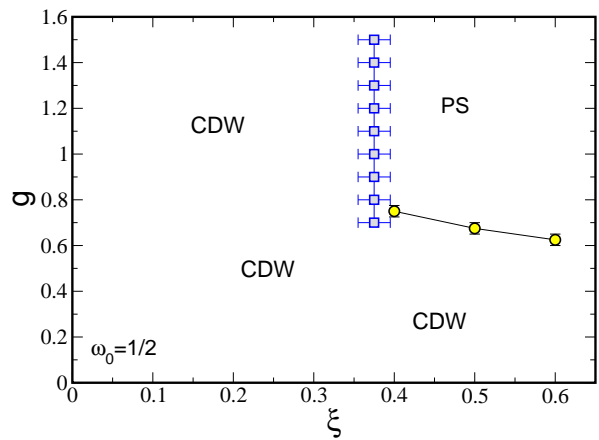}}
\caption{(Color online) The phase diagram in the $g$-$\xi$ plane at
  half-filling and constant $\beta=9.6$.  The phonon frequency is
  fixed at $\omega=1/2$.  At small $\xi$, the Holstein limit CDW
  correlations form.  This occurs even at weak coupling owing to the
  Fermi surface nesting on the square lattice.  As $\xi$ increases,
  phase separation occurs. }
\label{phasediag}
\end{figure}

Figure \ref{scdwvsxi} shows $S(\pi,\pi)/N$ for several lattice sizes
$N$ as a function of $\xi$ at fixed $\beta=9.6$, $\omega_0=0.5$ and
$\lambda_0=1.5$.  Data for different $N$ coincide at small $\xi$,
indicating long range order.  At $\xi \approx 0.4$, $S(\pi,\pi)$ falls
rapidly.  This parameter sweep represents one cut through the
phase diagram of Fig.~\ref{phasediag}.

\begin{figure}[!ht]
\centerline{\includegraphics[width=9cm]{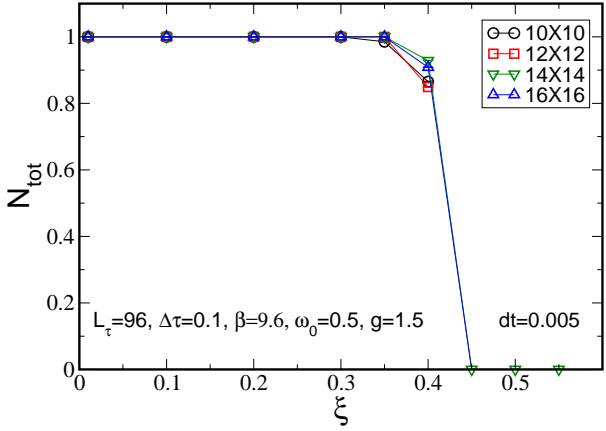}}
\caption{(Color online)  Same as Fig.~\ref{scdwvsxi} but showing the
  total number of particles. For $\xi \gtrsim 0.4$ the occupation
  vanishes, indicating phase separation. See Fig.~\ref{phasediag}. }
\label{ntotvsxi}
\end{figure}

Figure \ref{ntotvsxi}, which shows the density for the same parameters
as Fig. \ref{scdwvsxi}, suggests the collapse of CDW order is not due
to the density-density correlator becoming random (as would occur if
one heats the Holstein model above its $T_c$).  Instead, the particle
density plummets although the chemical potential is chosen to yield
half filling.  This suggests phase separation as $\xi$ grows.

\begin{figure}[!ht]
\centerline{\includegraphics[width=9cm]{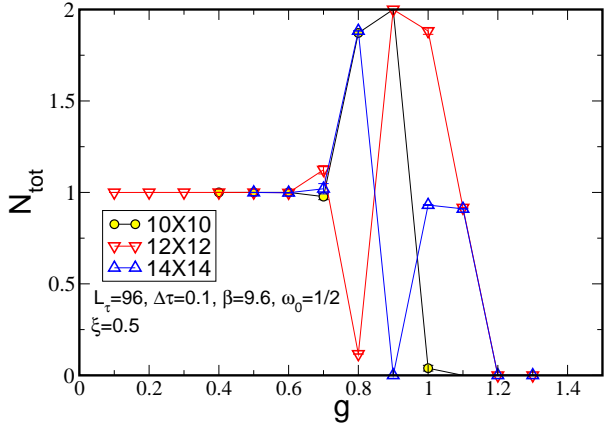}}
\caption{(Color online) The total occupation as a function of the
  interaction strength at fixed (large) $\xi$. For $\lambda_0 > 0.7$
  the system is either empty or totally full, indicating phase
  separation. See Fig. \ref{phasediag}.}
\label{ntotvslambda}
\end{figure}

Figure \ref{ntotvslambda} shows results in which instead $\lambda_0$
is increased at fixed $\xi=0.5$.  Large density fluctuations set in at
$\lambda_0 \gtrsim 0.7$, again indicating phase separation.  This
provides another cut in the $\lambda_0$-$\xi$ plane to generate the
phase boundary of Fig.~\ref{phasediag}.

\begin{figure}[!ht]
\centerline{\includegraphics[width=9cm]{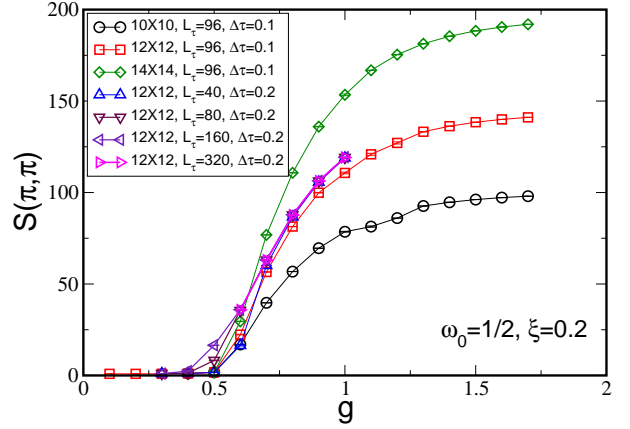}}
\caption{(Color online) $S(\pi,\pi)$ as a function of $\lambda_0$ at
  fixed $\xi =0.2$ where there is no phase separation (See
  Fig. \ref{phasediag}). As $\lambda_0$ decreases, $S(\pi,\pi)$
  decreases, becoming small for $\lambda_0 < 0.4$.  We believe this is
  due to an exponentially small $T_c$ which has decreased below our
  simulation temperatures $T \sim (1/16-1/64)\,t$ rather than a lack
  of CDW order.  See text.}
\label{scdwvslamba}
\end{figure}

\begin{figure}[!ht]
\centerline{\includegraphics[width=9cm]{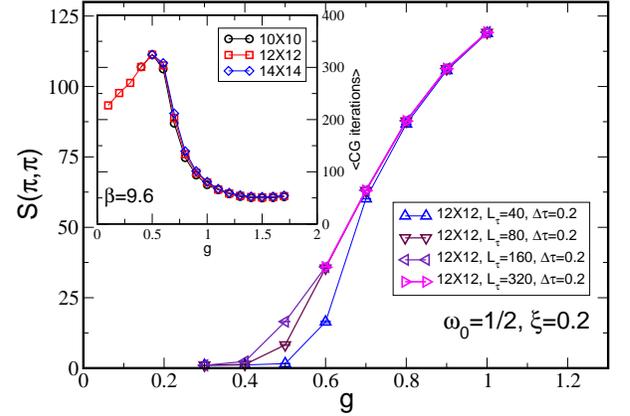}}
\caption{(Color online) Same as Fig.~\ref{scdwvslamba} but for one
  size and several values of $\beta$. $S(\pi,\pi)$ increases, and then
  saturates, as $\beta$ is increased. We are unable to see the CDW
  phase unambiguously for $\lambda_0 < 0.4$. Inset: For the same
  $\omega_0$ and $\xi$ and for $\beta=9.6$, we show the average number
  of CG iterations as a function of the coupling $g$. At weak
  coupling, the number of iterations is large, peaks at $g=0.5$, and
  decreases rapidly as the CDW gap strengthens. Note the absence of
  dependence on system size. }
\label{scdwvslambdabeta}
\end{figure}

Fixing $\xi=0.2$, $S(\pi,\pi)$ grows rapidly at $\lambda_0 \sim 0.6$
for $\beta=9.6$ (Fig.~\ref{scdwvslamba}).  We believe that the region
of small $S(\pi,\pi)$, $\lambda_0 \lesssim 0.6$, is associated with
the fact that the transition temperature is exponentially low.  We see
no evidence for phase separation.  This is also confirmed in
Fig.~\ref{scdwvslambdabeta} where one sees a growth of $S(\pi,\pi)$ in
the small $\lambda$ region as $T$ is lowered.  The observation of
ordered phases at weak coupling is often a subtle issue in finite
temperature simulations, since, for example, one often has BCS-like
functional forms $T_c \sim \omega e^{-ct/\lambda}$ which become
exponentially small as $\lambda$ decreases.  This is supported by the
fact that $S(\pi,\pi)$ grows as $\beta$ increases, unlike situations
where there is a disordered phase below a critical coupling value, as
occurs on a honeycomb lattice\cite{paiva05,zhang18}. The finite
temperature difficulty in observing the CDW phase was also noted in
Ref. \onlinecite{weber17}.

\begin{figure}[!ht]
\centerline{\includegraphics[width=9cm]{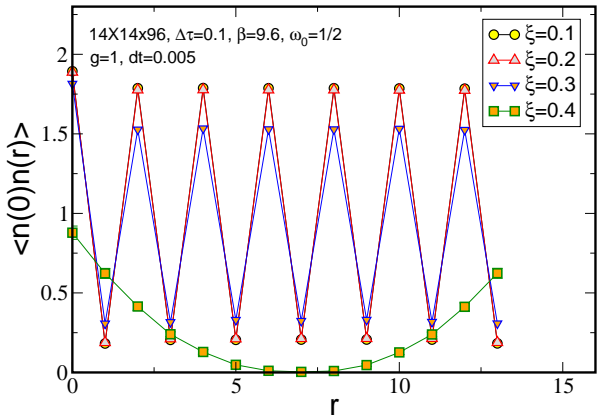}}
\caption{(Color online) The density correlation function as a function
  of distance at fixed $g=1$ and $\xi=0.1,\,0.2,\,0.3,\,0.4$. For the
  first three values of $\xi$, the system is in the CDW phase, as is
  clear from the robust oscillation of the correlation function. For
  $\xi=0.4$ the system has undergone phase separation.}
\label{denden}
\end{figure}

Although we have focused on the (momentum space) CDW structure factor,
one can also of course observe the oscillating density correlations
directly in real space.  An example is given in Fig.~\ref{denden}.  At
fixed $\beta=9.6$ and $g=1.0$, robust density oscillations occur up to
$\xi =0.3$. For $\xi=0.4$ the system has undergone phase separation
and the density correlation function no longer exhibits CDW
oscillations.

\begin{figure}[!ht]
\centerline{\includegraphics[width=9cm]{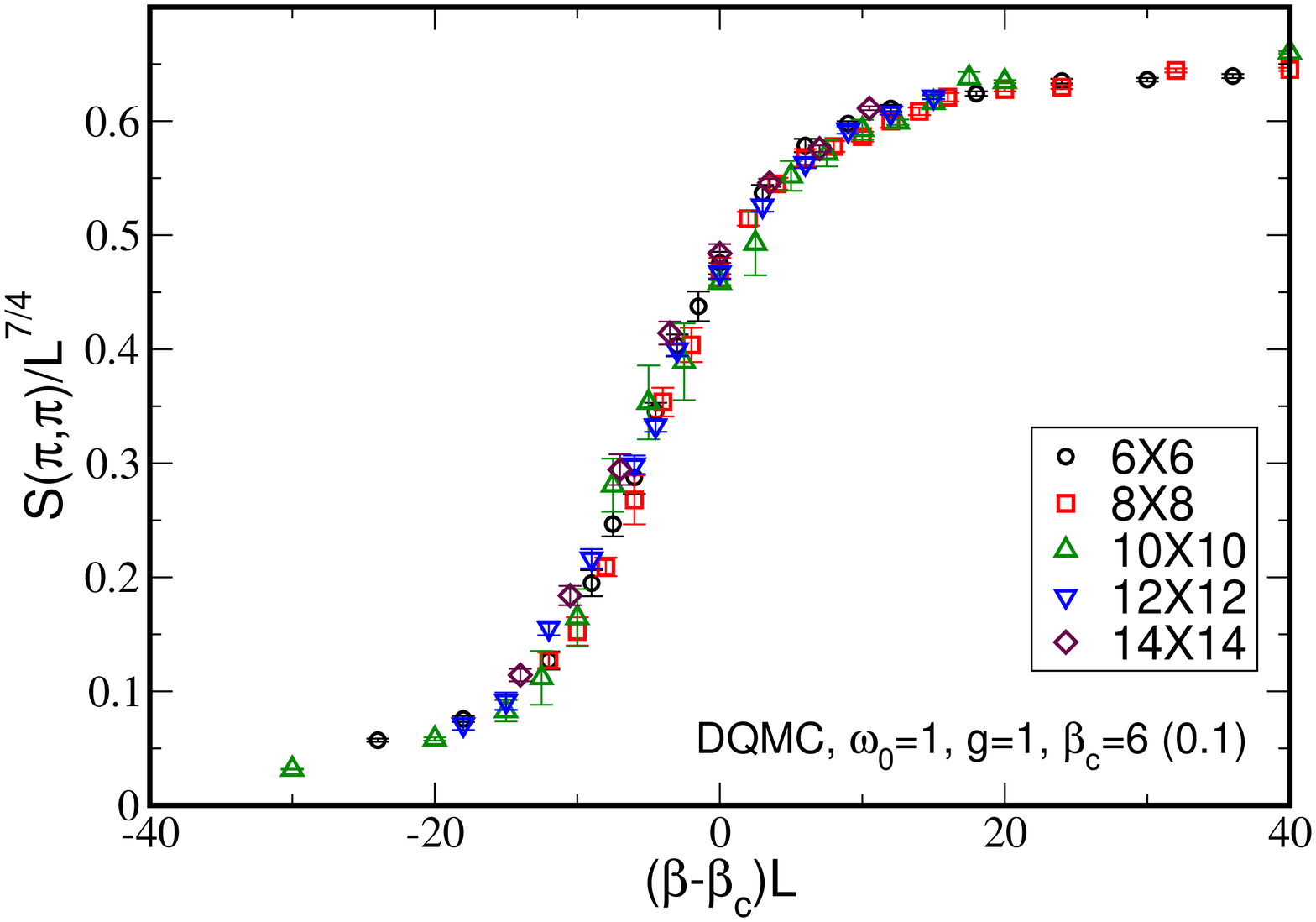}}
\centerline{\includegraphics[width=9cm]{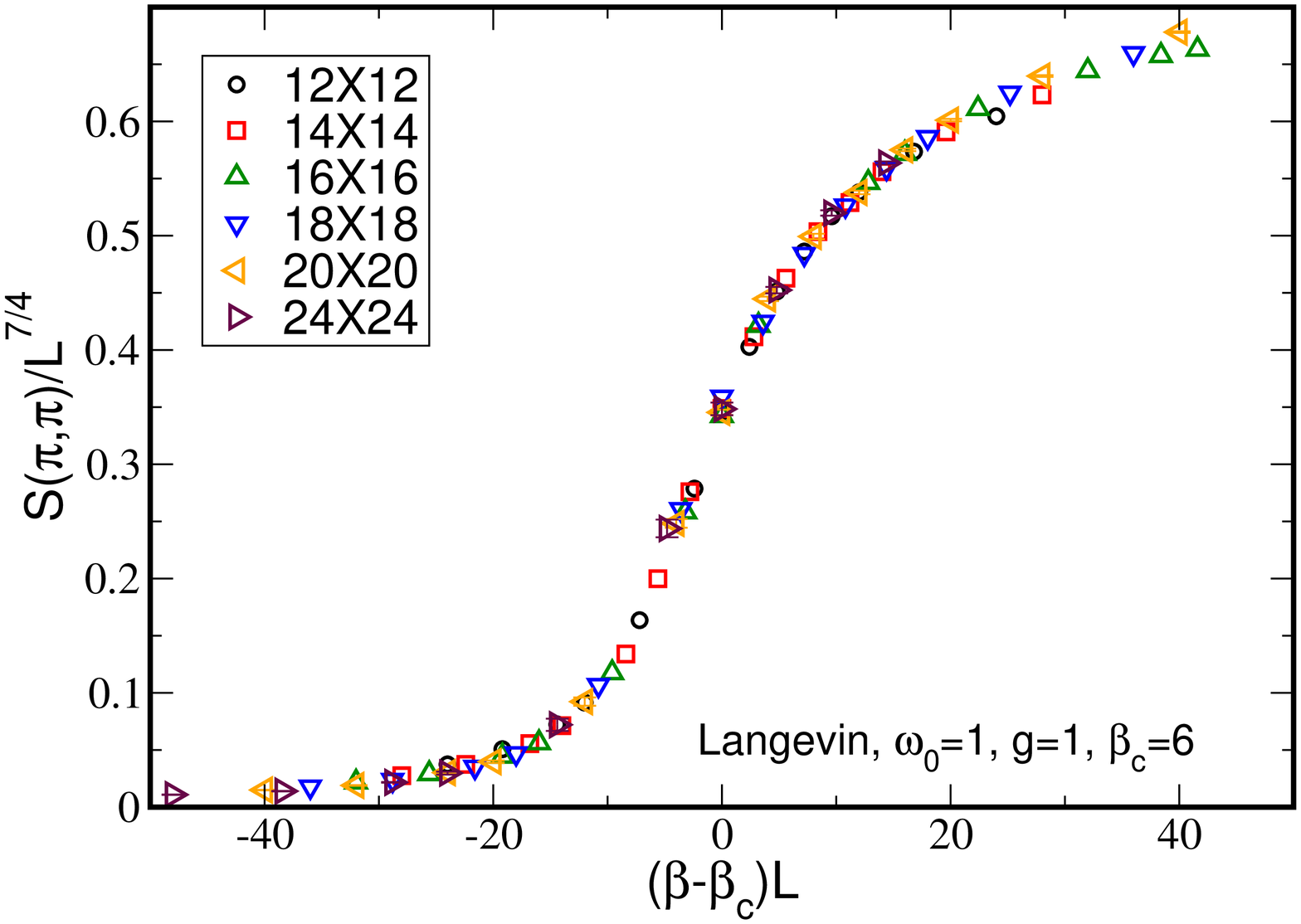}}
\caption{(Color online) Top: Scaling of the structure factor near the
  finite temperature transition from disordered to CDW
  phase\cite{costa18} using DQMC. Bottom: Same but using
  Langevin. Note the larger system sizes and the cleaner collapse in
  the latter case. In the lower panel, the error bars are smaller than
  the symbols.}
\label{langholsteinscaling}
\end{figure}

Taken together, Figs. \ref{scdwvsxi}-\ref{denden} can be used to
generate the phase diagram at $beta=9.6$ shown in
Fig.~\ref{phasediag}.  The small $\lambda$ regions are labeled as CDW
based on the earlier arguments concerning the critical temperature.

\begin{figure}[!ht]
\centerline{\includegraphics[width=9cm]{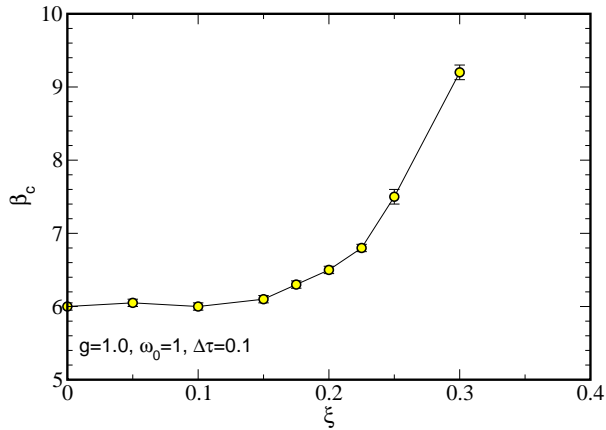}}
\caption{(Color online) The critical inverse temperature, $\beta_c$,
  as a function of the range of the electron-phonon interaction,
  $\xi$. $\beta_c$ was obtained by performing finite size scaling
  analysis as in Fig. \ref{langholsteinscaling}.}
\label{Tcvsxi}
\end{figure}

Figure \ref{phasediag} shows the phase diagram at very low
temperature.  Since CDW formation breaks a discrete symmetry, we
expect that in two dimensions the phase transition occurs at finite
critical temperarture, $T_c$.  Figure \ref{langholsteinscaling} shows
a finite size scaling analysis of $S(\pi,\pi)$, making use of the
expected Ising critical exponents\cite{scalettar89,noack91,weber17}.
The top panel uses data from the `traditional' DQMC approach which is
based on single phonon field updates and scales as ${\cal O}(N^3)$,
while the bottom panel shows the Langevin results with computation
time scaling linearly with $N$. The ability to simulate larger lattice
sizes is seen to provide a more convincing data collapse, which occurs
as the thermodynamic limit is approached. In addition, the efficiency
of the algorithm allows us to get more precise results in reasonable
computation time. As a benchmark, we note that the simulations of the
$12\times 12$ system took about four times longer with the Langevin
algorithm than with DQMC, but the precision is much better than a
factor of two. For larger systems sizes, the run time difference
favors Langevin even more because of the linear scaling as opposed to
cubic for DQMC. We performed such finite size scaling to find the
critical inverse temparature, $\beta_c(\xi)$, as $\xi$ is
increased. Figure \ref{Tcvsxi} shows the resulting phase diagram in
the ($\xi,\beta_c$) plane illustrating that as $\xi$ increases,
$\beta_c$ increases rapidly until phase separation occurs at large
enough $\xi_c \approx 0.55$. We did not determine $\beta_c$ beyond
$\xi=0.3$ since it becomes very large.

\section{Conclusions and Outlook} 

In this paper we have formulated a Langevin-based Quantum Monte Carlo
algorithm for an interacting electron-phonon Hamiltonian, augmented by
Fourier Acceleration.  Variation of the phonon field in the imaginary
time direction is moderated by the phonon kinetic energy.  As a
consequence, the rapid growth in the number of conjugate gradient
iterations needed in such approaches to the Hubbard model is not
present here.

We have presented tests of our method which quantify various
systematic errors.  Comparisons with single update ${\cal O}(N^3)$
simulations using DQMC show that the Langevin method gives physically
correct results, and also indicate that the cross-over where it
becomes more efficient occurs at lattice sizes $ N \sim 10^2$.

We have applied the method to an electron-phonon model with finite
range ({\it i.e.} momentum-dependent) interaction.  This is a natural
target, since single update algorithms scale as ${\cal O}(N^4)$ and
hence have proven extremely challenging.  The phase diagram obtained
indicates a strong tendency towards phase separation, even with
correlation lengths as small as $\xi \sim 0.35$, a situation in which
the near-neighbor coupling is nearly two orders of magnitude smaller
than the on-site interaction. 

This sensitivity to finite $\xi$ suggests that application of such
models to materials with momentum-dependent $\tilde \lambda(q)$ will
need to include some sort of electron-electron interaction to inhibit
phase separation.  This is a very challenging task owing to the
resulting sign problem of doped systems which would lead to a complex
Langevin. Such equations have been studied quite extensively in the
context LGT\cite{aarts17} and recently applied to the one dimensional
Hubbard model\cite{loheac17} and ultra-cold fermionic atoms with
unequal masses\cite{rammelmuller18}. The complex Langevin equation was
also recently applied to the Holstein-Hubbard model\cite{karakuzu18}
and shown to be very efficient in the parameter range
$U>g^2/\omega_0$.

More direct applications of our approach will be to the on-site
(Holstein) case, for which there is still an abundance of open
questions, including studies of 3D and layered 2D systems, accurate
determination of critical properties via finite size scaling, role of
anharmonicities, and high resolution of observables in momentum space,
all of which require large lattice sizes. In addition, SSH models
where the hopping parameter fluctuates due to the nuclear oscillations
can be treated efficiently with this algorithm. An interesting
question to address is the competition between the contact Holstein
coupling and the bond SSH coupling in determing the phase of the
system.

\section{Acknowledgements} 

The work of RTS was supported by DOE-SC0014671.  GGB is supported by
the French government, through the UCAJEDI Investments in the Future
project managed by the National Research Agency (ANR) with the
reference number ANR-15-IDEX-01. We thank Steven Johnston, WeiTing
Chiu, and Thomas Devereaux for useful conversations.

\end{document}